\documentclass[showpacs,twocolumn,prb]{revtex4}
\usepackage{graphicx}
\newcommand{\be}{\begin{equation}}
\newcommand{\ee}{\end{equation}}
\newcommand{\bea}{\begin{eqnarray}}
\newcommand{\eea}{\end{eqnarray}}
\newcommand{\bwt}{\begin{widetext}}
\newcommand{\ewt}{\end{widetext}}

\begin{document}
\title{Impact of a finite cut-off for the optical sum rule in the superconducting state}
\author{F. Marsiglio$^1$, E. van Heumen$^2$, and A.B. Kuzmenko$^2$}
\affiliation{$^1$Department of Physics, University of Alberta, Edmonton, Alberta,
Canada, T6G~2J1, \\
$^2$DPMC, Universit\'e de Gen\`{e}ve, 24 Quai
Ernest-Ansermet, CH-1211 Gen\`{e}ve 4, Switzerland}
\begin{abstract}
A single band optical sum rule derived by Kubo can reveal a novel kind of superconducting state. It relies, however,
on a knowledge of the single band contribution from zero to infinite frequency. A number of experiments over the
past five years have used this sum rule; their data has been interpreted in support of 'kinetic energy-driven superconductivity'. However, because of the presence of unwanted interband optical spectral weight, they necessarily
have to truncate their sum at a finite frequency. This work examines theoretical models where the impact of
this truncation can be examined first in the normal state, and then in the superconducting state.
The latter case is particularly important as previous considerations attributed the observed anomalous temperature dependence  as an artifact of a non-infinite cutoff frequency. We find that this is in fact not the case, and that the sign of the corrections from the use of a non-infinite cutoff is such that the observed temperature dependence is even more anomalous when proper account is taken of the cutoff. On the other hand, in these same models, we find that the strong observed temperature dependence in the normal state can be attributed to the effect of a non-infinite cutoff frequency.
\end{abstract}

\date{\today}
\maketitle

\section{introduction}

Kubo \cite{kubo57} formulated two optical sum rules: the first involves all
electrons in the system under study, and relates the total integrated area under
the real part of the optical conductivity to basic parameters, the electron charge,
the bare electron mass, and the electron density. The second sum rule focusses on a
single band near the Fermi level, and relates the integrated area associated with
intraband transitions to a single particle property:

\begin{equation}
W(T) \equiv \int_{0}^{+\infty }d\nu \mathop{\rm Re} \left[ \sigma_{\rm xx} (\nu
)\right] = {\pi e^{2} \over 4\hbar^2} \biggl\{ \frac{4}{N}\sum_{k}
{\partial^2\epsilon_k \over \partial k_x^2} n_{k}. \biggr\}%
\label{sumrule}
\end{equation}

Here, $n_k$ is the single electron occupation number, and $\epsilon_k$ is the
dispersion relation. Unlike the first Kubo sum rule, this relation depends
on the system particulars (like the band structure), and external parameters,
like the temperature (since $n_k$ varies with temperature).

Over the last decade a number of optical measurements have been made in
the high temperature superconductors \cite{basov99,molegraaf02,santander-syro03,ortolani05,carbone06b,vanheumen07}
which indicate an anomalous temperature dependence of the optical sum rule.
Two noteworthy observations have been made: first, in the normal state,
as the temperature is decreased, the optical sum, $W(T)$ increases. This is expected,
even in a model without interactions, because of the thermal factor in the occupation (Sommerfeld's expansion).
However, the observed increase is an order of magnitude higher than that expected from
a non-interacting model \cite{benfatto04}. Much of this discrepancy can be explained through interactions \cite{benfatto06,haule07} and/or phase fluctuations \cite{toschi05}. However, it is possible that more mundane explanations exist, as will be discussed below.

The second observation is that the optical spectral weight {\em increases} below
the superconducting transition temperature in a variety of optimally doped and underdoped high $T_c$ samples.
The 'standard' BCS result is that the
optical spectral weight should {\em decrease} below the superconducting transition
temperature \cite{marsiglio06a}. It is this second observation in particular \cite{molegraaf02}
that has captured much interest \cite{hirsch02}.
At present, the notion of 'kinetic energy-driven' superconductivity \cite{hirsch02}
is reasonably well supported by current optical sum rule measurements.

However, the optical sum requires integration over the entire spectral range. There are
technical difficulties with this program, many of which have been overcome \cite{kuzmenko07}.
In addition, there is a difficulty that the intraband contributions (required for the sum rule
quoted in Eq. (\ref{sumrule}) above) may not be so readily separated from the interband contributions. Experimentalists generally impose a cutoff in the frequency integration, and then
vary that cutoff to look for cutoff effects. If these are minimal, then one is satisfied that
the sum represents the intraband contributions alone, and hence the measurement is examining
the property described by Eq. (\ref{sumrule}). Almost immediately following the first ab-plane
sum rule measurements \cite{molegraaf02}, Karakozov et al. \cite{karakozov02} suggested that
both observations could be explained by the use of a finite cutoff frequency. They made estimates of the changes in the normal state due to a temperature dependent scattering rate caused by the electron-phonon interaction,
and in the superconducting state due to a temperature dependent superconducting order parameter. These estimates were generally ignored by other workers in the field, who instead relied on
the sum rule itself, and used the right hand side of Eq. (\ref{sumrule}) to study trends and
parameter dependencies. Recently Norman et al. \cite{norman07} have
investigated the cutoff effect in the normal state in considerable detail, and have found that
a cutoff frequency as used in the experiments can indeed lead to a substantial temperature
dependence of the optical sum in the normal state, in agreement with Karakozov et al. We
provide further support for this conclusion with calculations below. This means that the
strong temperature dependence observed in the normal state can be quantitatively understood with a weak coupling picture.

The primary contribution of this work is a model calculation of the effect of a finite
cutoff frequency in the superconducting state. We find that a finite cutoff leads to a decrease in the
optical spectral weight in the superconducting state (in a model where no change is expected).
%an
%{\em underestimate} of the expected 'standard' lowering of the optical spectral weight in the
%superconducting state.
This means that the observed {\em increase} in the superconducting state
represents a lower bound on the increase. This result is in the opposite direction of the estimate of Karakozov et al. {\cite{karakozov02}; their result was made for the dirty limit
(Mattis-Bardeen \cite{mattis58}) result, which turns out to have high frequency properties
that are very distinct from those with non-infinite scattering rate. The conclusion is that
the increase in the optical spectral weight in the superconducting state does indeed suggest a decrease in kinetic energy, or, in more conventional language, a collapse of electronic scattering, in the
superconducting state \cite{norman02,knigavko04,marsiglio06a}.

In this paper we first show results in the normal state, for a simple model of electrons
coupled to an Einstein boson mode. This leads to a temperature dependent scattering rate,
which, as first shown in Ref. (\onlinecite{karakozov02}), leads to significant temperature
dependence in the normal state, in qualitative agreement with the observations. The same temperature dependence has minimal effect at temperatures where the superconducting state sets in; the high frequency scattering rate is essentially unaffected by the onset of superconductivity. However, the temperature dependence of the order parameter will give rise to adjustments in the spectral weight at high frequency (as well as low), resulting in a potentially significant temperature
dependence in the optical spectral weight. This is the subject of the second part of the paper.

\section{normal state}

For the conductivity in the normal state we use the expression \cite{allen71}

\bea
&&\sigma(\nu + i\delta) = \nonumber \\
&&{\omega_p^2 \over 4 \pi} {i \over \nu} \int_{-\infty}^{+\infty}
{f(\omega - \nu) - f(\omega) \over \nu + i/\tau_{\rm imp} - \bigl[\Sigma(\omega+i\delta) +
\Sigma(\nu - \omega+i\delta)\bigr]},
\label{cond_norm}
\eea
where $\omega_p$ is the bare plasma frequency, $1/\tau_{\rm imp}$ is the electron-impurity
scattering rate (taken here to be independent of wave vector and frequency), $f(\omega) \equiv 1/(\exp{\beta \omega} + 1)$
is the Fermi function ($\beta \equiv 1/(k_B T)$, and $\Sigma(\omega+i\delta)$ is the self energy due to the electron-boson interaction \cite{remarkm1}. Note that we have not attempted vertex corrections; these have been discussed (see, for example, Ref. \onlinecite{allen04}), and are suspected to be small.
For definiteness, we use for the self energy the standard Migdal result, obtained for electron-phonon scattering \cite{allen82,marsiglio_chap}
\bea
\Sigma(z) = &&\int_0^\infty d \Omega \alpha^2F(\Omega) 
%\nonumber \\
\biggl[ \,
-2\pi i (N(\Omega) + {1 \over 2} ) + \nonumber \\
&&\psi \bigl( {1 \over 2} + i{\Omega - z \over 2\pi T} \bigr) -
\psi \bigl( {1 \over 2} - i{\Omega + z \over 2\pi T} \bigr)
\, \biggr],
\label{self}
\eea
where $N(\Omega) \equiv 1/(\exp{\beta \omega} - 1)$ is the Bose function, and
$\psi (x)$ is the digamma function, and the entire expression is required at $z = \omega + i \delta$.

The desired calculation is to use Eq. (\ref{cond_norm}) in the {\em partial} optical spectral weight,
\be
W(\nu_c) = {2 \over \pi} \int_0^{\nu_c} \ d\nu \ {\sigma_1(\nu) \over \omega_P^2/(4 \pi)},
\label{partial}
\ee
where we have included constants so that the integral is dimensionless, and, moreover, $W(\nu_c \rightarrow \infty) = 1$. It is important to note that with an infinite cutoff, i.e. in the
spirit of the single band Kubo sum rule, Eq. (\ref{sumrule}), Eq. (\ref{partial}) with the conductivity from Eq. (\ref{cond_norm}) inserted yields {\em no temperature dependence whatsoever}. This is because Eq. (\ref{cond_norm}) was derived with a band structure with quadratic dispersion, so Eq. (\ref{sumrule}) clearly indicates (on the right-hand-side) that a constant is expected. This is a good model to use then, because any temperature dependence observed from Eq. (\ref{partial}) can be definitely attributed to the cutoff.

Since the cutoff $\nu_c$ is often taken to be high compared to other energy scales in the
problem, it suffices to determine the optical conductivity accurately at high frequencies.
Karakozov et al. \cite{karakozov02} noted (see also Norman et al.\cite{norman07}) that at
{\em high} frequency the conductivity is, in fact, Drude-like. Indeed, one obtains
\begin{equation}
\sigma(\nu + i\delta) \approx {\omega_p^2/(4 \pi) \over 1/\tau_\infty - i \nu},
\label{cond_high}
\end{equation}
where
\begin{equation}
1/\tau_\infty(T) \equiv 1/\tau_{\rm imp} + 2 \pi \int_0^\infty \ d \Omega
\alpha^2F(\Omega) {\rm coth}\biggl({\beta \Omega \over 2}\biggr).
\label{onet_high}
\end{equation}

Then the partial sum rule expression (\ref{partial}) can be integrated analytically, and
one obtains
\be
W_{\rm Dr}(\omega_c) \approx (2/\pi){\rm tan}^{-1}\biggl({\omega_c \over 1/\tau_\infty(T)}\biggr).
\label{approx}
\ee
How accurate is the expression in Eq. (\ref{approx}) compared with that obtained by integrating the full expression in Eq. (\ref{cond_norm}) ?
To answer this question and examine trends we use an Einstein boson with
frequency $\omega_E$ and electron-boson coupling strength $\lambda$. Then $\alpha^2F(\Omega)
= {\lambda \omega_E \over 2} \delta (\Omega - \omega_E)$, and the integral required
in the self energy can be done analytically. The frequency scale
$\omega_E$ represents the mean frequency of a broader $\alpha^2F(\Omega)$ spectrum, such as
those used by Norman et al. \cite{norman07}.

In Fig. 1 we show the real part of the conductivity vs. frequency for a system of electrons
interacting with impurities ($1/\tau_{\rm imp} = 10$ meV) and with Einstein bosons ($\lambda = 1$, and
$\omega_E = 60$ meV). In the low frequency range (part (a)), the Drude approximation does not agree at all with the more precise Kubo result, particularly at low temperatures. However, in the high frequency range (part (b)), the Drude approximation agrees very well with
the Kubo result, particularly beyond 2 eV. In Fig. 2 we show similar results
for an example system with parameters identical to those in Fig. 1, except
that $\omega_E = 20$ meV. It is clear that for lower boson frequency (same $\lambda$),
the Drude approximation works very well for even lower frequencies.

Note that here we have purposefully used a high frequency scattering rate to reproduce well
the {\em high} frequency part of the conductivity. As we remarked above, this is required for the partial optical sum. When an understanding of the low frequency
conductivity is required, a low frequency Drude form can be obtained through a different
approximation. Details are given in Ref. \onlinecite{marsiglio95}.

\begin{figure}[tp]
\begin{center}
\includegraphics[height=4.9in,width=3.5in]{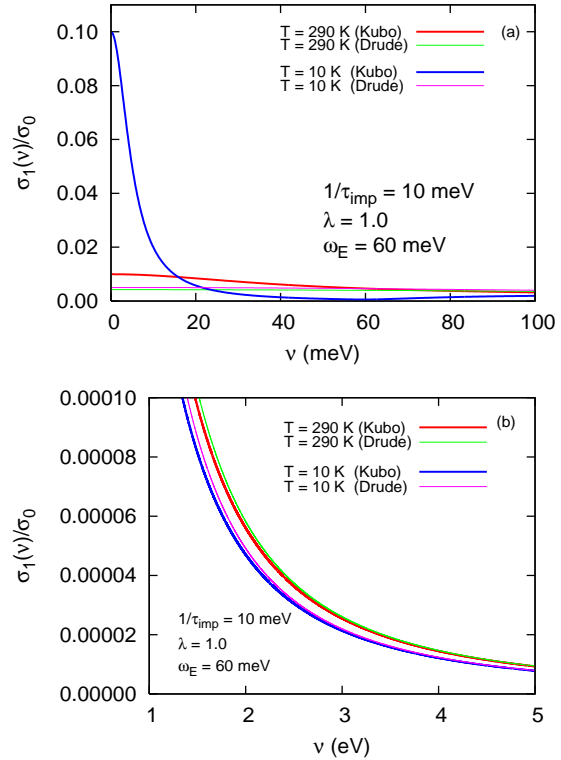}
\caption{ (color online) The real part of the optical conductivity (normalized to $\sigma_0 \equiv {\omega_p^2
\over 4 \pi} \tau$) vs. frequency, at two different temperatures. In either case we use the
full Kubo formulation (labelled 'Kubo' --- see Eq. (\protect\ref{sumrule})) and the so-called 'Drude' result (labelled 'Drude' --- see Eq. (\protect\ref{approx})), which is designed to be accurate at high frequency. In (a) we show the low frequency part, where discrepancies can be large, particularly at low temperatures. In (b) we show the high frequency tail, where the 'Drude' result, based on the high frequency scattering rate (Eq. (\protect\ref{onet_high})) is very accurate. This high frequency accuracy means that the conductivity (and hence the spectral weight) beyond some frequency will be well represented by the simpler Drude expression. Here, $\nu \approx 2$ eV suffices for high accuracy.}
\end{center}
\end{figure}
\begin{figure}[tp]
\begin{center}
\includegraphics[height=4.9in,width=3.5in]{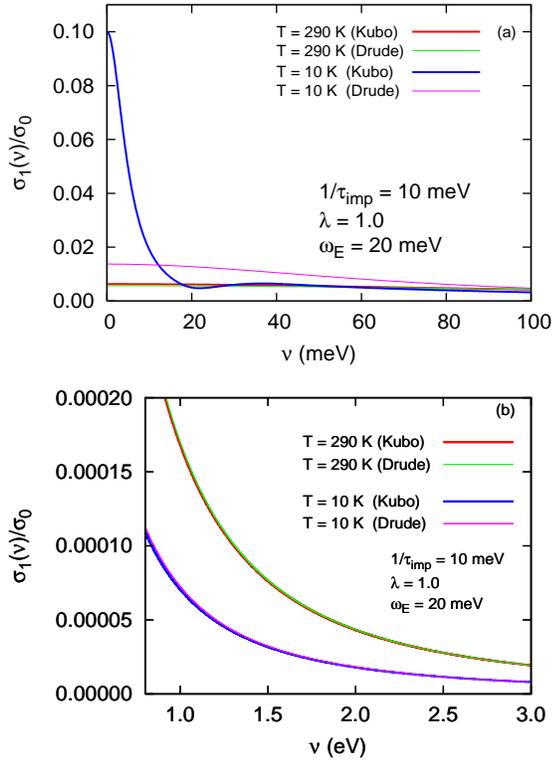}
\caption{ (color online) As in Fig. 1, the real part of the optical
conductivity (normalized to $\sigma_0 \equiv {\omega_p^2
\over 4 \pi} \tau$) vs. frequency, at two different temperatures. Now the Einstein boson frequency is 20 meV, compared with that in Fig. 1, which was 60 meV. The legend and curve designation are as in Fig. 1.
%For the two different temperatures we use the full Kubo formulation (labelled 'Kubo' --- see Eq. (\protect\ref{sumrule})) and the so-called 'Drude' result (labelled 'Drude' --- see Eq. (\protect\ref{approx})), which is designed to be accurate at high frequency. In (a) we show the low frequency part, where discrepancies can be large, particularly at low temperatures. In (b) we show the high frequency tail, where the 'Drude' result, based on the high
%frequency scattering rate (Eq. (\protect\ref{onet_high})) is very accurate, basically for all frequencies shown.
Comparison with Fig. 1  illustrates that for lower boson frequency (i.e. overall lower effective high frequency scattering rate), the Drude approximation is more precise for lower frequencies.}
\end{center}
\end{figure}

To examine the temperature dependence in the normal state, we show in Fig. 3 the optical spectral weight up to some frequency $\nu_c$ (= 1 eV in this case)
vs. temperature for several model systems. Note that in {\em all} cases the integral shown approaches unity for $\nu_c \rightarrow \infty$.
Several key points are evident in this figure. First, in the absence of inelastic scattering,
($\lambda = 0$, uppermost curve), there is no temperature dependence. The full sum rule (corresponding to $W(\nu_c = \infty) = 1$ with our normalization) is not quite achieved, due to the non-zero elastic scattering rate (we use $1/\tau_{\rm imp} = 10$ meV for all these results). For a fixed boson frequency, $\omega_E = 40$ meV, we increase the electron-boson coupling, $\lambda$ (top three curves). It is clear that (a) the overall weight decreases with increasing coupling, and (b) the temperature dependence also increases. Moreover, the Drude approximation (shown with symbols) is less accurate as boson coupling increases. All of this can be understood through Eq. (\ref{onet_high}), as the effective scattering rate increases with increasing coupling.

\begin{figure}[tp]
\begin{center}
\includegraphics[height=4.0in,width=3.5in]{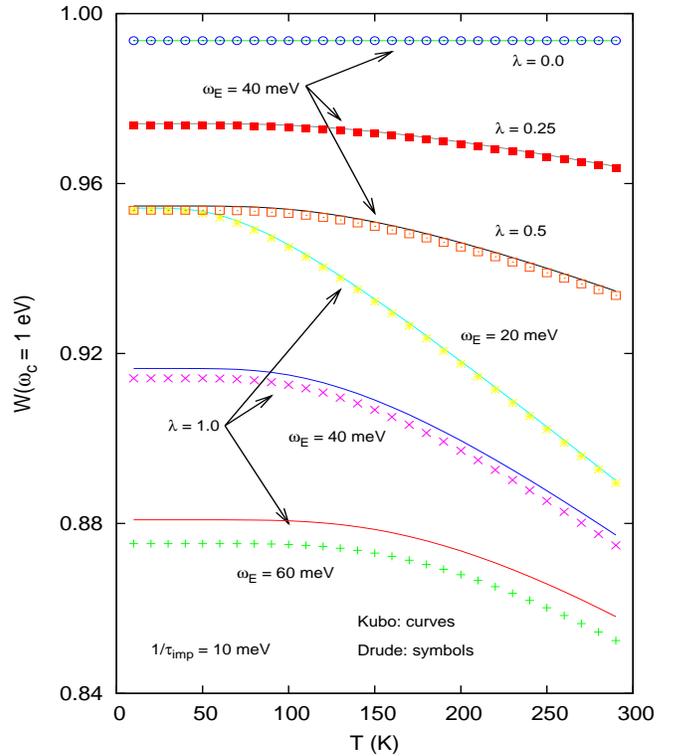}
\caption{ (color online) The optical spectral weight, integrated up to 1 eV, vs. temperature, for
a variety of model parameters.
Curves are for results using the Kubo formula, Eq. (\protect\ref{sumrule}), while accompanying
symbols are for the Drude approximation, given analytically in Eq. (\protect\ref{approx}). The lowest three curves and symbols explore the dependence with boson frequency (fixed coupling strength), and show that, in the temperature range of interest, increasing the boson frequency results in less temperature dependence. Since increasing the boson frequency also increases
the effective (infinite frequency) scattering rate (see Eq. (\protect\ref{onet_high})), the sum rule up to 1 eV is lower for higher boson frequency.
Starting from the 2nd lowest curve (indicated to have $\omega_E= 40$ meV), the top three
curves explore the dependence on coupling strength. Clearly as coupling strength decreases
the temperature dependence diminishes, until, with $\lambda = 0$ and therefore only elastic
scattering remaining, there is no temperature dependence (top curve). That two of the
curves merge at low temperature is not a coincidence; they have the same $\lambda \omega_E$
product, and therefore have the same low temperature sum rule result (see Eq. (\protect\ref{onet_high}) with an Einstein spectrum substituted for $\alpha^2F(\Omega)$).
}
\end{center}
\end{figure}

For fixed coupling strength, the effect of boson frequency is also shown in Fig. 3.
Clearly, as the boson frequency increases (lower three curves), the sum rule is less fulfilled, the temperature dependence diminishes, and the Drude formula becomes less accurate. The decreased temperature dependence is due to a scaling with temperature from Eq. (\ref{onet_high}): $1/\tau_\infty = 1/\tau_{\rm imp} + \pi \lambda \omega_E {\rm coth}\bigl(\beta \omega_E /2\bigr)$. Less accuracy is achieved with the Drude formula for higher boson frequency because the 'high' frequency limit for the Drude formula is no longer achieved at 1 eV when the boson frequency becomes higher than about 50 meV. The actual temperature dependence of the partial sum rule is exponential (for an Einstein boson). However, when Fig. 3 is re-plotted vs. $T^2$, linear behaviour is observed for many of the parameters, over the temperature range relevant
for the normal state (say, above 90 K). Norman et al. \cite{norman07} also found remarkable agreement with a $T^2$ temperature dependence, though it was not explicit in their case either. In any event no further attempt is made here to optimize the agreement with experiment; we have no doubt that this could be done, but would require adjustment of unknown boson spectral functions and unknown band structure parameters.

The key point (made by Norman et al.\cite{norman07}) is that the magnitude of change
in the normal state can be very large (of order 2\% or more). The origin of this variation with temperature is the finite cutoff. Clearly this variation easily surpasses the amount seen experimentally \cite{molegraaf02,santander-syro03,ortolani05,carbone06b,vanheumen07}, and so the first observation alluded to in the introduction may merely be due to a finite cutoff in the optical spectral sum.

\section{superconducting state}

The question then arises, can a similar cutoff effect, in the same direction (i.e. increase with decrease in temperature), occur in the superconducting state, in which case the observed anomalous behaviour could also be attributed to a finite cutoff ? The answer is {\em no}, as we now explain.

First, note that the integrations must be done with care, as the effects we are trying to discern are as little as one tenth of a percent. Hence, we will focus on zero temperature,
where some of the integrals involved can be done more accurately. We define
\be
\Delta W(\nu_c) \equiv W_S(\nu_c) - W_N(\nu_c),
\label{diff}
\ee
where the subscripts refer to the superconducting and normal states, respectively. Again, for the models discussed here, this quantity is zero as $\nu_c \rightarrow \infty$, as the right hand side of Eq. (\ref{sumrule}) is a constant.

First, how does the effective infinite frequency scattering rate change in the superconducting
state? The answer is that it does not at all. As Kaplan et al. \cite{kaplan76} showed, the scattering rate in the superconducting state is significantly more complicated than in the normal state. However, as might be expected, in the high frequency limit, this expression reduces to that in the normal state, given by Eq. (\ref{onet_high}). Hence one
can imagine that, just as in the normal state, the important frequency dependence in the superconducting state will be
encapsulated in the infinite frequency limit of $1/\tau(\omega)$. For this reason, for the rest of this paper we
focus on the BCS limit of Eliashberg theory, where the order parameter is
not frequency-dependent. Another reason for doing this is that it more clearly separates the two questions; the first,
the impact of the temperature dependence of the scattering on the sum rule, has already been addressed by normal state
calculations in the first part of the paper. For the second question, the impact of the temperature dependence of the order parameter, a BCS calculation allows us to focus only on this aspect, and does not include any remnant temperature dependence of the scattering rate \cite{remark0}.

Nonetheless, some caution may be in order. Inspection of Fig. 5 (below) best illustrates the problem; on the scale
of this figure, one cannot tell
whether the superconducting and normal state conductivities actually cross. It may be that some subtle refinement of
the theory may alter this state of affairs; for example, everything discussed here applies for an order parameter with
s-wave symmetry. To our knowledge no one has performed these difficult calculations with a d-wave order parameter. This
will be the subject of future work.

% fig. 4
\begin{figure}[tp]
\begin{center}
\includegraphics[height=4.0in, width=3.5in]{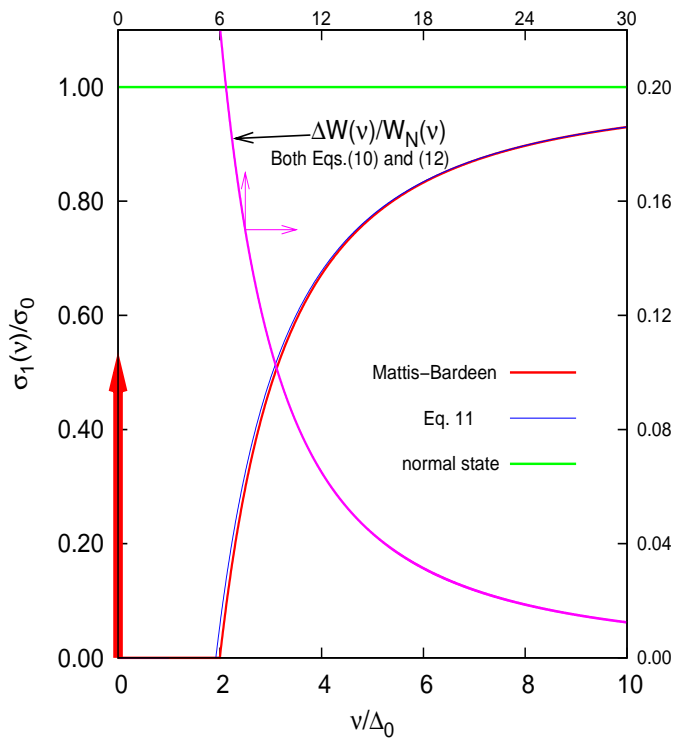}
\caption{ (color online) The Mattis-Bardeen result for the optical conductivity (left scale) and the optical spectral
weight (right scale), as a function of frequency (note the bottom scale is for the conductivity and
the top scale is for the spectral weight). In the Mattis-Bardeen limit the normal state is a constant; this is indicated
by the horizontal (green) line at unity. The full numerical result for the conductivity is labelled and shown by the solid (red) curve, including a delta-function contribution at the origin. The analytical result, given by Eq. (\protect\ref{jahnke}), is also shown, and is in remarkable agreement with the numerical result right down to $\nu = 2\Delta_0$. The full numerical result for the optical spectral weight integrated up to a frequency $\nu$ is shown
by the solid (pink) curve. The approximate result, Eq. (\protect\ref{diff2_approx}), is indiscernible from the numerical
result. Note that as the frequency increases, the Mattis-Bardeen result always lies beneath the normal state.}
\end{center}
\end{figure}

\subsection{Mattis-Bardeen limit}

Following Karakozov et al. \cite{karakozov02} we first make another simplification --- we use the Mattis-Bardeen (MB), or
dirty limit, where analytical expressions (at $T=0$) are available.
We have \cite{mattis58}
\be
{\sigma_S(\nu) \over \sigma_0} = {\pi^2 \over 4} \delta(\bar{\nu}) + \theta(\bar{\nu} - 1)
\biggl( {1 + \bar{\nu} \over \bar{\nu}} E(m) - {2 \over \bar{\nu}} K(m) \biggr)
\label{mattisbardeen58}
\ee
where $\bar{\nu} \equiv {\nu \over 2 \Delta_0}$, with $\Delta_0$ the zero temperature
energy gap, and $m \equiv [(\bar{\nu} - 1)/(\bar{\nu} + 1)]^2$, with the complete elliptic integral of the first and
second kind defined \cite{abramowitz64}
$K(m) \equiv \int_0^{\pi/2} \ d\theta \ {1 \over (1 - m {\rm sin}^2{\theta})^{1/2}}$, and
$E(m) \equiv \int_0^{\pi/2} \ d\theta \ (1 - m {\rm sin}^2{\theta})^{1/2}$, respectively.
Here $\sigma_0$ is the zero frequency normal state conductivity.
This expression is easily integrated up to some finite cutoff; however, we require the
difference in the optical spectral weight defined by Eq. (\ref{diff}), so the high
frequency part is of most interest. Since $\Delta W(\infty) = 0$, we have
\be
\Delta W(\nu_c) = - \int_{\nu_c}^\infty \ d \nu \ (\sigma_S(\nu) - \sigma_N(\nu)),
\label{diff2}
\ee
and only the high frequency part need be obtained accurately in Eq. (\ref{mattisbardeen58}).

Alternatively, for the Mattis-Bardeen result, we find, for high frequency \cite{jahnke45}
\be
{\sigma_S(\nu) \over \sigma_0} \approx 1 - \biggl( {\Delta_0 \over \nu} \biggr)^2 \bigl[ 1 + 2 \log{{2 \nu \over \Delta_0}}
\bigr].
\label{jahnke}
\ee
Then the integral in Eq. (\ref{diff2}) can be done analytically, and the result is
\be
{\Delta W(\nu_c) \over W_N(\nu_c)} \approx \biggl({\Delta_0 \over \nu_c}\biggr)^2 \bigl[3 + 2
\log{{2 \nu_c \over \Delta_0}}\bigr].
\label{diff2_approx}
\ee
Note that $W_N(\nu_c) = \sigma_0 \nu_c$.
Both the conductivity and the optical spectral weight difference as defined by Eq. (\ref{diff2}) are plotted in Fig. 4 (the latter is normalized to the normal state optical integral), along with their approximate counterparts, Eq. (\ref{jahnke}) and Eq. (\ref{diff2_approx}), respectively. Remarkably, the approximation for the conductivity
is barely distinguishable from the full result, all the way down to $\nu \sim 2 \Delta_0$,
while the approximation for the sum rule difference cannot be seen on this plot (it is beneath
the numerical result). Note the difference in both horizontal and vertical scales for the two
quantities.
%Since $\Delta_0$ is anticipated to be about 30 meV or so, we have shown the normalized sum rule difference $\Delta W(\nu_c)$
%out to about $30 \times$ the gap energy, or close to 1 eV.
The (normalized) sum rule difference is still more than 1\% at $30 \times$ the gap energy (top horizontal scale). This indicates that, because of the non-infinite frequency cut off, the optical spectral weight
is expected to {\em increase} by a 'small' amount. The experiments reporting an anomalous increase in the superconducting state in the a-b axis conductivity \cite{molegraaf02,santander-syro03,carbone06b,vanheumen07}
typically report a change of less than 1/2\%, so the 'small amount' referred to above is
really not so small.

% fig. 5
\begin{figure}[tp]
\begin{center}
\includegraphics[height=3.2in,width=3.0in, angle = -90]{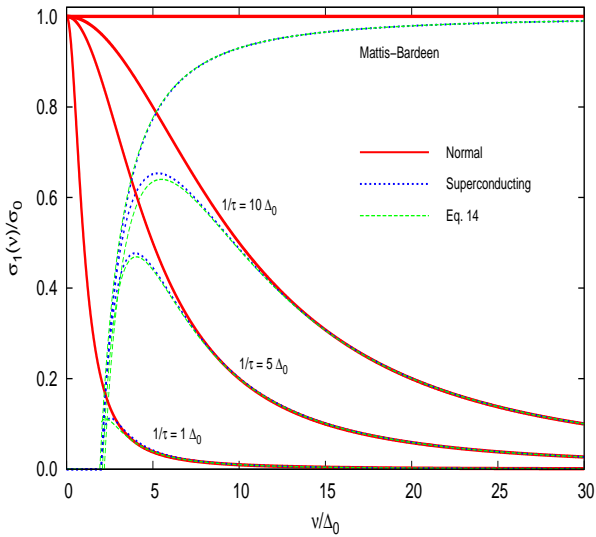}
\caption{(color online) The real part of the optical conductivity (normalized to
$\sigma_0 \equiv {\omega_p^2 \over 4 \pi} \tau$)
vs. frequency, for a variety of scattering rates. Solid red curves are in the normal state,
and are simple Drude results. The dotted blue curves are the results of the numerical integration in the superconducting
state; they all have a gap at $\nu = 2 \Delta_0$. The high frequency expansions for the superconducting state (Eq. (\protect\ref{approx1})) are shown by the dashed green curves. These are surprisingly very accurate right down to $\nu = 2 \Delta_0$. The Mattis-Bardeen result is also shown --- it is quite obvious that the result in the superconducting state is
always beneath that in the normal state. For the other scattering rates shown, however, the superconducting state crosses
the normal state, a characteristic first noted by Chubukov et al. \protect\cite{chubukov03}, and emphasized in Fig. 6 below.}
\end{center}
\end{figure}

If this was the entire story, one could conclude that all 'anomalous' observations, i.e.
the strong temperature dependence in the normal state, and the anomalous increase in the
superconducting state, can be attributed to the use of a non-infinite frequency cutoff,
as suggested by Karakozov and coworkers \cite{karakozov02}. Then these experimental results would {\em not} indicate any novel kind of physics. However, as we now illustrate, the Mattis-Bardeen limit is pathological in this matter, and the optical sum rule in the
case of a non-infinite scattering rate behaves qualitatively very differently than the MB limit.

\subsection{Non-infinite scattering rate}

To our knowledge, Chubukov et al. \cite{chubukov03} first noticed that, within a BCS formalism, the conductivity
in the superconducting state actually exceeds that in the normal state. For large scattering rates they found that the crossover occurred at a frequency comparable to the scattering rate, $1/\tau$. This then implies that the optical sum, obtained by integrating out to some high frequency, will become {\em lower} in the superconducting state than in the normal state.

% fig. 6
\begin{figure}[tp]
\begin{center}
\includegraphics[height=3.5in,width=3.0in, angle = -90]{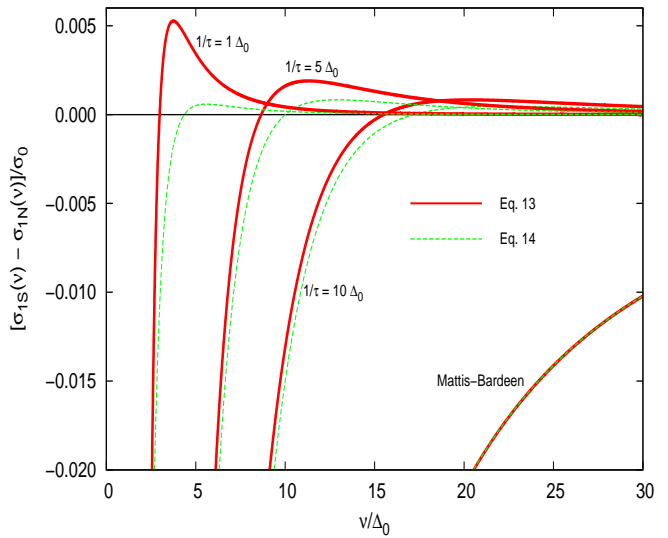}
\caption{(color online) The difference (superconducting - normal) in the real part of the optical conductivity (normalized to $\sigma_0 \equiv {\omega_p^2 \over 4 \pi} \tau$) vs. frequency, for the cases shown in Fig.~5.
Solid red curves are for the numerical result, and green curves are for the analytical high frequency result given by
Eq. (\protect\ref{approx1}). The accuracy is remarkable (note the vertical scale). In particular the frequency at which the
difference crosses zero is given by the approximate expression fairly accurately. Note that the Mattis-Bardeen approximate result (Eq. (\protect\ref{jahnke})) is invisible under the numerical result.}
\end{center}
\end{figure}

Thus for any non-infinite scattering rate, a finite frequency cutoff (presumed to be higher than the scattering rate) will result in a lowering of the optical sum in the superconducting state.
Recall, that in the models we use here, provided we integrate to infinite frequency, there should be no change in the optical sum rule as a function of temperature, even in the superconducting state. Thus, the observed increase actually underestimates the 'true' amount, since experiments are unable to integrate the optical spectral weight out to infinite frequency.

To illustrate this, we again confine ourselves to zero temperature, where integrals, etc. can
be done more precisely. A somewhat compact expression for the real part of the conductivity, for any scattering rate is given by \cite{lee89,zimmermann91,marsiglio91a}
\bea
\sigma_1(\nu) = &&{ne^2 \over m} {1 \over 2\nu} {\rm Im} \int^{\nu - \Delta_0}_{\Delta_0}
\ d \omega \nonumber \\
& & \biggl\{ {1+N(\omega) N(\nu - \omega) - P(\omega) P(\nu - \omega) \over \epsilon (\nu - \omega) + \epsilon(\omega) - i/\tau}  \nonumber \\
- & & {1-N(\omega) N(\nu - \omega) + P(\omega) P(\nu - \omega) \over \epsilon (\nu - \omega) - \epsilon(\omega) - i/\tau} \biggr\},
\label{sigma1}
\eea
where we have used $\sigma_0 = {ne^2 \tau \over m}$. Note that $\epsilon(\omega) \equiv \sqrt{\omega^2 - \Delta_0^2}$, and $N(\omega) = \omega/\epsilon(\omega)$ and  $P(\omega) = \Delta_0/\epsilon(\omega)$, and all quantities in Eq. (\ref{sigma1}) are real except for the
explicit imaginary $i/\tau$ in the denominators. No special definition for the square-roots is
required since the frequency $\omega$ is always positive \cite{remark1}.

Results from Eq. (\ref{sigma1}) are obtained numerically. However, for large frequencies, an expansion is possible, and we
obtain, to second order in $\Delta_0/\nu$, for any value of $1/\tau$, \cite{remark2}
\bea
{\sigma_{1S}(\nu) \over \sigma_0} \approx &&{ (1/\tau)^2 \over \nu^2 + (1/\tau)^2} \biggl(
1 - 2 \bigl({\Delta_0 \over \nu}\bigr)^2 \bigl[1 + \log{2 \nu \over \Delta_0} \bigr] \nonumber \\
&& \phantom{aaa} - 2 { \Delta_0^2 \over \nu^2 + (1/\tau)^2} \bigl[1 - 2\log{2 \nu \over \Delta_0} \bigr] \biggr).
\label{approx1}
\eea

% fig. 7
\begin{figure}[tp]
\begin{center}
\includegraphics[height=3.5in,width=3.0in, angle = -90]{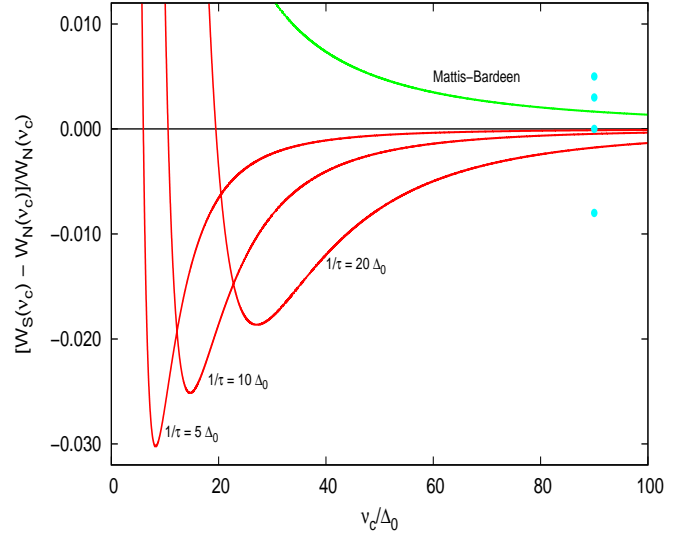}
\caption{(color online) The normalized difference in the optical spectral weight for a variety of scattering rates, $1/\tau$.
Note that for any non-infinite scattering rate, this difference is negative for frequencies above approximately
$1/\tau$. A negative difference has the same sign as the conventional BCS result. The experimental results for various doping levels in Bi$_2$Sr$_2$CaCu$_2$O$_{8+\delta}$ \protect\cite{carbone06b} are shown by the points, with a cutoff frequency
$\nu_c = 1.24$ eV and assumed gap of $14$ meV. Note that the underdoped and optimally doped results are positive. }
\end{center}
\end{figure}

These results (dashed green curves) are plotted along with the numerical results (dotted blue curves) in Fig.~5 for
a variety of scattering rates. The high frequency expansion given by Eq. (\ref{approx1}) is remarkably accurate over
the entire frequency range. Also shown is the Mattis-Bardeen limit (see Eq. (\ref{jahnke}) above) which was encountered
already in Fig.~4. For the present discussion, the important characteristic of Eq. (\ref{approx1}) is that it crosses
the normal state result (solid red curves) at a frequency close to that given by the numerical result. Note that in Fig.~5 a possible crossing is not even apparent; hence in Fig.~6 we plot the difference in conductivities vs. frequency. Once again solid red (dashed green) curves
refer to the full numerical (approximate analytical (Eq. (\ref{approx1}))) result. Here we see that the frequency $\nu_x$, at which the difference in the conductivities ($\sigma_{1S} - \sigma_{1N}$) crosses zero, is reasonably well described by the asymptotic formula above (dashed green curves) compared with the numerical results (solid red curves). From inspection of the figure, as the scattering rate increases, $\nu_x$ also increases, and of course the approximation for this improves, since we used a high frequency expansion. Note how well the {\em difference} in the Mattis-Bardeen limit is described by Eq. (\ref{approx1}) (or Eq. (\ref{jahnke})). Analytically, the crossover frequency is given (approximately) by the solution
to $\nu_x = {1 \over \tau} \sqrt{1+1/\log{u_x} \over 1 - 2/\log{u_x}}$, where $u_x \equiv 2\nu_x/\Delta_0$. This crossover frequency approaches (very slowly) $1/\tau$ as $1/\tau \rightarrow \infty$.

The important point, however, is that the result as seen in Fig.~6 (and not really apparent in Fig.~5) shows a qualitatively different behaviour for the difference in conductivities when a non-infinite scattering rate is present. At frequencies somewhat higher than the scattering rate, the conductivity in the superconducting state always overshoots, by a small amount, the conductivity in the normal state \cite{chubukov03}. This feature is not present in the Mattis-Bardeen limit, and the latter is misleading on this point.

It should now be apparent that an integration of the conductivity in the superconducting state, say, up to a frequency of
about $30 \Delta_0$, will yield a small spectral weight compared to that in the normal state (since the curves in Fig.~6
are greater than zero above $30 \Delta_0$, and the total spectral weight of the difference up to infinite frequency has to integrate to zero, this statement follows). In Fig.~7 we plot the normalized difference in spectral weights (superconducting minus normal) up to a cutoff frequency $\nu_c$. As anticipated, for finite scattering rate the expected optical sum difference up to some cutoff $\nu_c$, is always less than zero. This means that the 'error' incurred by integrating up to some non-infinite frequency will reinforce the conventional \cite{marsiglio06b} BCS result that leads to a prediction that the
optical sum difference is negative. The data measured for four different doping levels is indicated by the symbols. Our results indicate that the observed positive change (in underdoped and optimally doped samples) may in fact be even slightly larger once the finite cutoff is accounted for. The degree of this correction depends on the scattering rate, which is not known with any certainty. For example, with an assumed scattering rate of $20 \Delta_0$, and $\Delta_0 \approx 15 - 25$ meV,
this yields a value of $1/\tau \approx 300 - 500$ meV. This appears to be quite high, but Eq. (\ref{onet_high}) indicates that for a boson mode with coupling strength $\lambda = 1$ at $\Omega_E = 50$ meV, one obtains $1/\tau \approx 300$. Thus values in this range can be expected.

\section{Summary}

The main conclusion of this work is that the single band Kubo sum rule, even with the caveat that
an infinite frequency cutoff is not possible due to experimental limitations, remains an important diagnostic of
novel superconductivity. In particular the notion \cite{hirsch02} and observation \cite{molegraaf02,santander-syro03,carbone06b,vanheumen07} of 'kinetic energy-driven' superconductivity remain valid
in spite of the non-infinite frequency cutoff limitation. The key point in the theory is that the high frequency behaviour
of the conductivity (in the superconducting vs. the normal state) with a non-infinite scattering rate is qualitatively different than that in the Mattis-Bardeen (dirty) limit \cite{chubukov03}. In the latter case a finite cutoff always
causes an increase in the optical spectral weight in the superconducting state compared to the normal state, whereas in
the more realistic case of a non-infinite scattering rate, the opposite is true.

\begin{acknowledgments}

This work was supported in part by the
Natural Sciences and Engineering Research Council of Canada (NSERC),
by ICORE (Alberta), and by the Canadian Institute for Advanced Research
(CIfAR). The work at the University of Geneva is supported by the Swiss National Science Foundation
through grant 200020-113293 and the National Center of Competence in Research (NCCR) Materials
with Novel Electronic Properties-MaNEP. FM is grateful to the Aspen Center for
Physics, where much of this work was done. Discussions with Andrey Chubukov and Mike Norman are gratefully acknowledged.
We wish to thank Dirk van der Marel for both suggesting this calculation and encouragement throughout this work.
\end{acknowledgments}

\bibliographystyle{prb}

\begin{thebibliography}{1}
\bibitem{remarkm1} We use the phrase electron-boson, since a large literature in this field
has used the Migdal formalism adopted here for bosons other than phonons. The general applicability
of this formalism is still an open question, but we expect it is accurate for the electron-phonon
interaction: see Ref. \onlinecite{allen04}.
\bibitem{remark0} Inspection of Fig. 3 shows that with a superconducting transition temperature of 80 K,
there are two possibilities. The first is exemplified by the lowest curve shown, where, essentially there
is no temperature dependence below $T_c$. Then as we have argued above, we expect no other temperature
dependence in the superconducting state {\em from the scattering rate}, and only temperature dependence from
the opening of a superconducting gap will affect the optical spectral weight. It is with this case in mind
that we have argued that the BCS limit of Eliashberg theory is appropriate. However, even in the second case,
exemplified by the curve in Fig. 3 with $\omega_E = 20$ meV, where the normal state is still changing as
a function of temperature at $T_c$, we expect that BCS theory will best capture the effect of the opening of
a gap, because the relevant experimental quantity is the difference at $T=0$ of the {\em observed}
spectral weight in the superconducting state and the {\em extrapolated} spectral weight in the normal state.
This extrapolation procedure already accounts for additional temperature dependence from the scattering rate that
may be present below $T_c$; we expect this will be similar in the superconducting and normal states.
\bibitem{remark1} An earlier contour integration requires that  ${\rm sgn} \bigl({\rm Re} \sqrt{\omega^2 -
\Delta_0^2}\bigr)={\rm sgn}(\omega)$. This makes quantities like the density of states $N(\omega)$ even functions of $\omega$. In this case we have kept all these arguments positive, so there is no special condition required in the final expression.
\bibitem{remark2} Note that the result in Ref. \onlinecite{chubukov03} (see their Eq. 9) agrees with ours to logarithmic accuracy; we include other terms of order $(\Delta_0/\nu)^2$ so that our agreement with the numerical result is considerably better.
\end{thebibliography}

\end{document}